\documentclass[prd,twocolumn,amsmath,amssymb]{revtex4}
\usepackage{graphicx}

\begin{document}

\title{\boldmath
Direct Measurement of the Pseudoscalar Decay
Constant $f_{D^+}$}
\author{
\begin{small}
M.~Ablikim$^{1}$,              J.~Z.~Bai$^{1}$,               Y.~Ban$^{11}$,
J.~G.~Bian$^{1}$,              X.~Cai$^{1}$,                  J.~F.~Chang$^{1}$,
H.~F.~Chen$^{16}$,             H.~S.~Chen$^{1}$,              H.~X.~Chen$^{1}$,
J.~C.~Chen$^{1}$
% ...
\end{small}
\\(BES Collaboration)\\
}
\vspace{0.2cm}
\affiliation{
\begin{minipage}{145mm}
$^{1}$ Institute of High Energy Physics, Beijing 100049, People's Republic of China\\
% ...
$^{11}$ Peking University, Beijing 100871, People's Republic of China\\
% ...
$^{16}$ University of Science and Technology of China, Hefei 230026, People's Republic of China\\
% ...
\vspace{0.4cm}
\end{minipage}
}

\begin{abstract}    
The absolute branching fraction for the decay
$D^+ \rightarrow \mu^+ \nu$ has been directly measured based on
a data sample of about 33 ${\rm pb^{-1}}$ 
collected around $\sqrt{s}=3.773$ GeV
with the BES-II detector at the BEPC collider.
A total of $5321 \pm 149 \pm 160$ $D^-$ mesons are reconstructed 
in nine hadronic decay modes.
In the system recoiling against these singly tagged $D^-$ mesons, $2.67\pm1.74$
purely leptonic decay events of $D^+ \rightarrow \mu^+ \nu_{\mu}$ are observed.
Those yield the branching fraction of
$BF(D^+ \rightarrow \mu^+ \nu_{\mu}) = (0.122^{+0.111}_{-0.053}\pm 0.010)\%$,
and a corresponding value of the pseudoscalar decay constant
$f_{D^+}=(371^{+129}_{-119}\pm 25)$ MeV.
\end{abstract}

\maketitle

\section{\bf Introduction} 
   In the standard model, the $D^+$ (Through this
Letter, charge conjugation is implied) meson can decay into
$l^+\nu_l$ (where $l^+$ is $e^+$, $\mu^+$ or $\tau^+$)
through a virtual $W^+$ boson. The virtual $W^+$ boson is produced
in the annihilation of the $c$ and $\overline d$ quarks.
The decay rate of this process is determined by the wavefunction overlap of
the two quarks at the origin, and is parametrized by the $D^+$ decay
constant, $f_{D^+}$.
Fig. 1 shows the decay diagram for the 
Cabibbo-suppressed purely leptonic decay of the $D^+$ meson.
The decay width of the $D^+ \rightarrow l^+\nu_l$ 
is given by the formula~\cite{Rosner}
{\bf
\begin{center}
\begin{equation}
\Gamma(D^+ \rightarrow l^+\nu_{l})=
     \frac{G^2_F f^2_{D^+}} {8\pi}
      \mid V_{cd} \mid^2
      m^2_l m_{D^+}
    \left (1- \frac{m^2_l}{m^2_{D^+}}\right )^2,
\end{equation}
\end{center}
}
\noindent
\noindent
where $G_F$ is the Fermi coupling constant, $V_{cd}$ is the 
$c\rightarrow d$ Cabibbo-Kobayashi-Maskawa (CKM) matrix 
element~\cite{pdg}, $m_l$ is the mass of the lepton, and
$m_{D^+}$ is the mass of the $D^+$ meson.
\begin{figure}[hbt]
\includegraphics[width=7.0cm,height=3.5cm]
{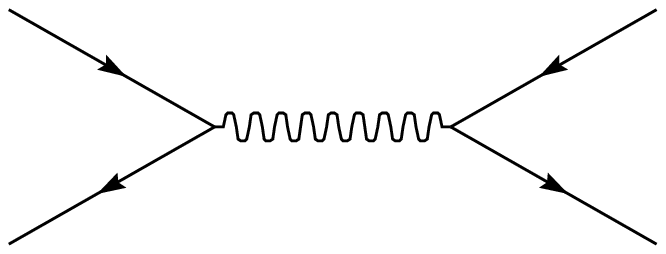}
\put(-222,45){\large\bf $D^+$}
\put(-12,80){\large\bf $l^+$}
\put(-12,10){\large\bf $\nu_l$}
\put(-195,80){\large\bf $c$}
\put(-195,10){\large\bf $\bar d$}
\put(-105,60){\large\bf $W^+$}
\caption{The decay diagram for $D^+ \rightarrow l^+\nu_l$.
}
\end{figure}

$f_{D^+}$ is an important parameter. 
However, it is difficult to measure $f_{D^+}$ due to the fact 
that the $D^+ \rightarrow l^+\nu_l$ is a
Cabibbo-suppressed decay process. 
There are some theoretical calculations to
estimate the value of $f_{D^+}$. Although predictions for $f_{D^+}$ vary
significantly from 90 to 360 MeV~\cite{prdct_fd},
the predictions for the ratios $f_{D^+}:f_{D^+_S}:f_{B}$ are more
reliable, where $f_{D^+_S}$ and $f_{B}$ are the decay constants
for the $D^+_S$ and the $B$ mesons, respectively.
The meson decay constants play an important role in extracting some
interesting physics quantities from diverse measurements. For example,
$f_{B}$ relates the measurement of the 
$B \bar B$
mixing~\cite{bbmix}
ratio to CKM matrix elements. At present it is not
possible to determine $f_{B}$ experimentally
from the purely leptonic $B$ decay, so theoretical calculations of
$f_{B}$ must be used. Hence, the calculations of $f_{B}$
are of considerable importance.
With the predictions for the
ratios $f_{D^+}:f_{D^+_S}:f_{B}$, measurements of $f_{D^+}$
and $f_{D^+_S}$ provide checks on some theoretical calculations of the decay
constants and help discriminate among different models and improve the
reliability of estimates of $f_{B}$.

   To date, there are eight experimental measurements of $f_{D^+_S}$ from the
WA75~\cite{wa75}, CLEO~\cite{cleo}, BES~\cite{bes_fds},
E653~\cite{e653}, L3~\cite{l3}, BEATRICE~\cite{beatrice}, OPAL~\cite{opal}
and ALEPH~\cite{aleph} groups.
For $D^+ \rightarrow \mu^+ \nu_{\mu}$,
the MARK-III Collaboration~\cite{markiii} set a 
branching fraction upper limit of $0.07\%$
(corresponding to $f_{D^+}<290$ MeV at $90\%$ C.L.).
The BES Collaboration~\cite{bes_fd} measured
$f_{D^+} = (300^{+180+80}_{-150-40})$ MeV based on one event of
$D^+ \rightarrow \mu^+ \nu_{\mu}$ from the data collected at 4.03 GeV with
the BES-I detector at the BEPC collider.

  In this Letter, we report a direct measurement of the branching
fraction for the decay $D^+ \rightarrow \mu^+ \nu_{\mu}$ and 
determination of the decay constant $f_{D^+}$.

\section{\bf The BES-II DETECTOR}
The BES-II is a conventional cylindrical magnetic detector that is
described in detail in Ref.~\cite{bes}.  A 12-layer vertex chamber
(VC) surrounding the beryllium beam pipe provides input to the event
trigger, as well as coordinate information.  A forty-layer main drift
chamber (MDC) located just outside the VC yields precise measurements
of charged particle trajectories with a solid angle coverage of $85\%$
of $4\pi$; it also provides ionization energy loss ($dE/dx$)
measurements which are used for particle identification.  Momentum
resolution of $1.7\%\sqrt{1+p^2}$ ($p$ in GeV/c) and $dE/dx$
resolution of $8.5\%$ for Bhabha scattering electrons are obtained for
the data taken at $\sqrt{s}=3.773$ GeV. An array of 48 scintillation
counters surrounding the MDC measures the time of flight (TOF) of
charged particles with a resolution of about 180 ps for electrons.
Outside the TOF, a 12 radiation length, lead-gas barrel shower counter
(BSC), operating in self-quenching streamer mode,
measures the energies of
electrons and photons over $80\%$ of the total solid angle with an
energy resolution of $\sigma_E/E=0.22/\sqrt{E}$ ($E$ in GeV) and spatial
resolutions of
$\sigma_{\phi}=7.9$ mrad and $\sigma_Z=2.3$ cm for
electrons. A solenoidal magnet outside the BSC provides a 0.4 T
magnetic field in the central tracking region of the detector. Three
double-layer muon counters instrument the magnet flux return, and serve
to identify muons of momentum greater than 500 MeV/c. They cover
$68\%$ of the total solid angle.

\section{DATA ANALYSIS}

The data used in the
analysis were collected with the BES-II detector at the BEPC collider.
A total integrated luminosity of about 33 $\rm pb^{-1}$ was taken at and
around the center-of-mass energy of 3.773 GeV. Those are
just above the threshold of $e^+e^- \rightarrow D\overline D$
and below the threshold of $e^+e^- \rightarrow D\overline D^*$.
Thus, if a $D^-$ meson decay is fully reconstructed 
(This is called a singly tagged $D^-$ meson),
a $D^+$ meson must exist in the system recoiling against the
singly tagged $D^-$ meson.
From the singly tagged $D^-$ event sample,
the events of the decay $D^+ \rightarrow \mu^+ \nu_{\mu}$ 
can be well selected in the recoiling system. 
Therefore, the absolute branching fraction for
the decay $D^+ \rightarrow \mu^+ \nu_{\mu}$
can be well measured and the decay constant $f_{D^+}$ can be determined.

\subsection{Singly tagged $D^-$ event sample}

\subsubsection{Events selection}

The $D^-$ meson is reconstructed in the
nine hadronic decay modes of 
$K^+\pi^-\pi^-$, $K^0\pi^-$, $K^0K^-$, $K^+K^-\pi^-$,
$K^0\pi^-\pi^-\pi^+$, $K^0\pi^-\pi^0$,  $K^+\pi^-\pi^-\pi^0$,
$K^+\pi^+\pi^-\pi^-\pi^-$ and $\pi^+\pi^-\pi^-$.
Events which contain at least
three reconstructed charged tracks with good helix fits are selected.
In order to ensure
the well-measured 3-momentum vectors and the reliably charged particle
identification, the charged tracks used in the single tag analysis
are required to be within $|cos\theta|<$0.85, 
where $\theta$ is the polar angle of the charged track.
All tracks, save those from $K^0_S$ decays, must originate
from the interaction region,
which require that the closest approach of the charged track
in the $xy$ plane is less than 2.0 cm and
the absolute $z$ position of the track is less than 20.0 cm.
Pions and kaons are identified by means of
TOF and $dE/dx$ measurements. Pion identification requires a consistency
with the pion hypothesis at a confidence level ($CL_{\pi}$) greater than
$0.1\%$.
In order to reduce misidentification, a kaon candidate is
required to have a larger confidence level ($CL_{K}$) for a kaon hypothesis
than that for a pion hypothesis.
The $\pi^0$ is reconstructed in the decay of
$\pi^0 \rightarrow \gamma\gamma$.
To select good photons from the decay
of $\pi^0$, the energy of a photon deposited in the BSC
is required to be greater than $0.07$ GeV~\cite{smlp1_pbl}, and the electromagnetic shower
is required to start in the first 5 readout layers. In order to reduce
backgrounds, the angle between the
photon and the nearest charged track is required to be greater
than $22^{\circ}$~\cite{smlp1_pbl}
and the angle between the direction of the cluster development
and the direction of the photon emission to be less than
$37^{\circ}$~\cite{smlp1_pbl}.

For the single tag modes of $D^- \rightarrow K^+\pi^+\pi^-\pi^-\pi^-$ and
$D^- \rightarrow \pi^+\pi^-\pi^-$, backgrounds are further reduced by
requiring the difference between the measured energy
of the $D^-$ candidate and the beam energy 
to be less than 70 and 60 MeV, respectively.
In addition, the cosine of the $D^-$ production angle relative to the beam
direction is required to be $|cos\theta_{D^-}|<0.8$.

\subsubsection{Single tag analysis}

For each event, there may be several different charged track (or
charged and neutral track) combinations for each of
the nine single tag modes.
Each combination is subject to a one-constraint (1C) kinematic fit requiring
overall event energy conservation and that the unmeasured recoil system has
the same invariant mass as the track combinations.
Candidates with a fit probability $P(\chi^2)$ greater
than $0.1\%$ are retained.
If more than one combination 
satisfies
$P(\chi^2)>0.1\%$,
the combination with the largest fit probability is retained.
For the single tag modes
with a neutral kaon and/or neutral pion,
one additional constraint kinematic fit
for the $K^0_S \rightarrow \pi^+\pi^-$ and/or
$\pi^0 \rightarrow \gamma\gamma$ hypothesis is performed, separately.

\begin{figure}[hbt]
\includegraphics[width=9.0cm,height=11.5cm]
{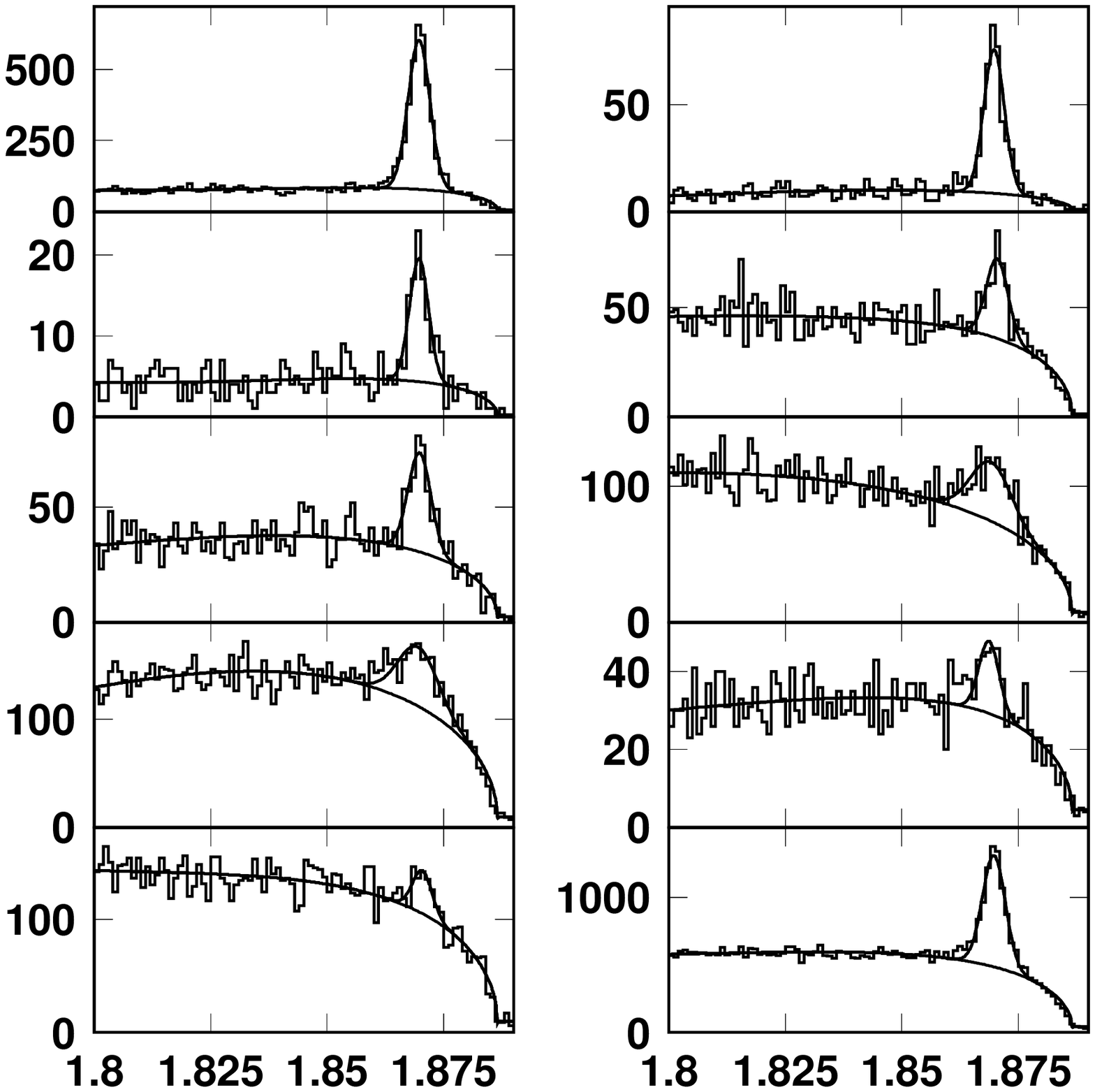}
\put(-165,5){Invariant Mass (GeV/$c^2$)}
\put(-260,130){\rotatebox{90}{Events/(0.001 GeV/$c^2$)}}
\put(-205,295){(a)}
\put(-90,295){(b)}
\put(-205,245){(c)}
\put(-90,245){(d)}
\put(-205,155){(e)}
\put(-90,155){(f)}
\put(-205,105){(g)}
\put(-90,105){(h)}
\put(-205,50){(i)}
\put(-90,50){(j)}
\caption{
The distributions of the fitted invariant masses of 
(a) $K^+\pi^-\pi^-$, 
(b) $K^0\pi^-$, 
(c) $K^0K^-$, 
(d) $K^+K^-\pi^-$,
(e) $K^0\pi^-\pi^-\pi^+$, 
(f) $K^0\pi^-\pi^0$,
(g) $K^+\pi^-\pi^-\pi^0$, 
(h) $K^+\pi^+\pi^-\pi^-\pi^-$, 
and
(i) $\pi^-\pi^-\pi^+$ combinations;
(j) the 9 modes combined together; the curves are the best fits described in
the text.
}
\label{sgltg}
\end{figure}

The resulting distributions in the fitted invariant masses
of $mKn\pi$ 
($m=0~ {\rm or}~1~{\rm or}~2$ and $n=1~{\rm or}~2~{\rm or}~3~{\rm or}~4$) 
combinations,
which are calculated using the fitted momentum vectors from the kinematic
fit, are shown in Fig.~\ref{sgltg}.
The signals for the singly tagged $D^-$ mesons are clearly observed
in the fitted mass spectra.
A maximum likelihood fit
to the mass spectrum with a Gaussian function for the $D^-$
signal and a special
background function\footnote{
A Gaussian function was assumed for the signal. The
background shape was
$$(1.0+p_1 y + p_2 y^2)
N\sqrt{1-(\frac{x}{E_b})^2}~x~e^{-f(1-\frac{x}{E_b})^2}+c,$$
where $N\sqrt{1-(\frac{x}{E_b})^2}~x~e^{-f(1-\frac{x}{E_b})^2}$
is the ARGUS background shape,
$x$ is the fitted mass, $E_b$ is the beam energy,
$y=(E_b-x)/(E_b -1.8)$,
$N$, $f$, $p_1$, $p_2$ and c are the fit parameters.
The parameter c accounts for the varying of the beam energy.
The ARGUS background shape was used by the ARGUS experiment to parameterize
the
background for fitting $B$ mass peaks. For details,
see~\cite{mnfit}.
}
to describe backgrounds
yields the number of the singly tagged $D^-$ events
for each of the nine modes and the total number of $5321\pm 149 \pm 160$
reconstructed $D^-$ mesons, where the
first error is statistical and the second systematic obtained
by varying the parameterization of the background.
The curves of Fig.~\ref{sgltg} give the best fits to the invariant mass
spectra. In the fits to the mass spectra, the standard deviations of the
Gaussian signal functions for Fig.~\ref{sgltg}(g) and
Fig.~\ref{sgltg}(h) 
are fixed at 4.27 MeV and 2.16 MeV, respectively. 
These standard deviations are obtained from Monte Carlo
sample. All other parameters are left free in the fit.

\subsection{Events of $D^+ \rightarrow \mu^+ \nu_{\mu}$}
\subsubsection{Muon Identification}

To identify the muon from the decay $D^+\rightarrow \mu^+\nu_{\mu}$,
the charged track in the recoil system of the $D^-$ tag in each of the
events as shown in Fig.~\ref{sgltg}
is examined for muon identification requirements.
The muon identification requires:
  \begin{enumerate}
    \item
     The charged track satisfies $|\cos \theta| < 0.68$,
     where $\theta$ is the polar angle of the charged track.

    \item
     There must be hits in the muon system and
     the hits must well associate
     with the charged track extrapolated from the track
     reconstructed in the MDC.
     The required number of the hits in the muon system
     is momentum dependent.

    \item
     The muon candidate with the transverse momentum of greater than
     0.7 GeV/$c$ is required to hit at least two layers 
     of the muon system.
  \end{enumerate}
\noindent
A charged track satisfying the 3 requirements is identified as a muon.
However, a small fraction of the pions which can punch through the muon system
could be misidentified as muons.
 
To estimate the average probability of misidentifying a pion as a muon
with momentum between 0.785 and 1.135 GeV/$c$,
we select the pions from the decay $J/\psi \rightarrow \omega \pi^+\pi^-$
in the data taken at the center-of-mass energy
of 3.097 GeV with the BES-II detector.
The pion 
which satisfies $|cos\theta_{\pi}|<0.68$ ($\theta_{\pi}$ 
is the polar angle of the pion)
and with momentum in the region from 0.785 to 1.135 GeV/$c$
is checked for whether it satisfies the muon selection requirements.
A total of 10657 pions from the $J/\psi \rightarrow \omega \pi^+\pi^-$ events 
satisfy the pion selection criteria and 259 of them are
misidentified as muons. Those give the averaged misidentification
probability of $0.024 \pm 0.002$.

\subsubsection{Candidates for $D^+ \rightarrow \mu^+ \nu_{\mu}$}

\begin{figure}[hbt]
\includegraphics[width=8.5cm,height=7.5cm]
{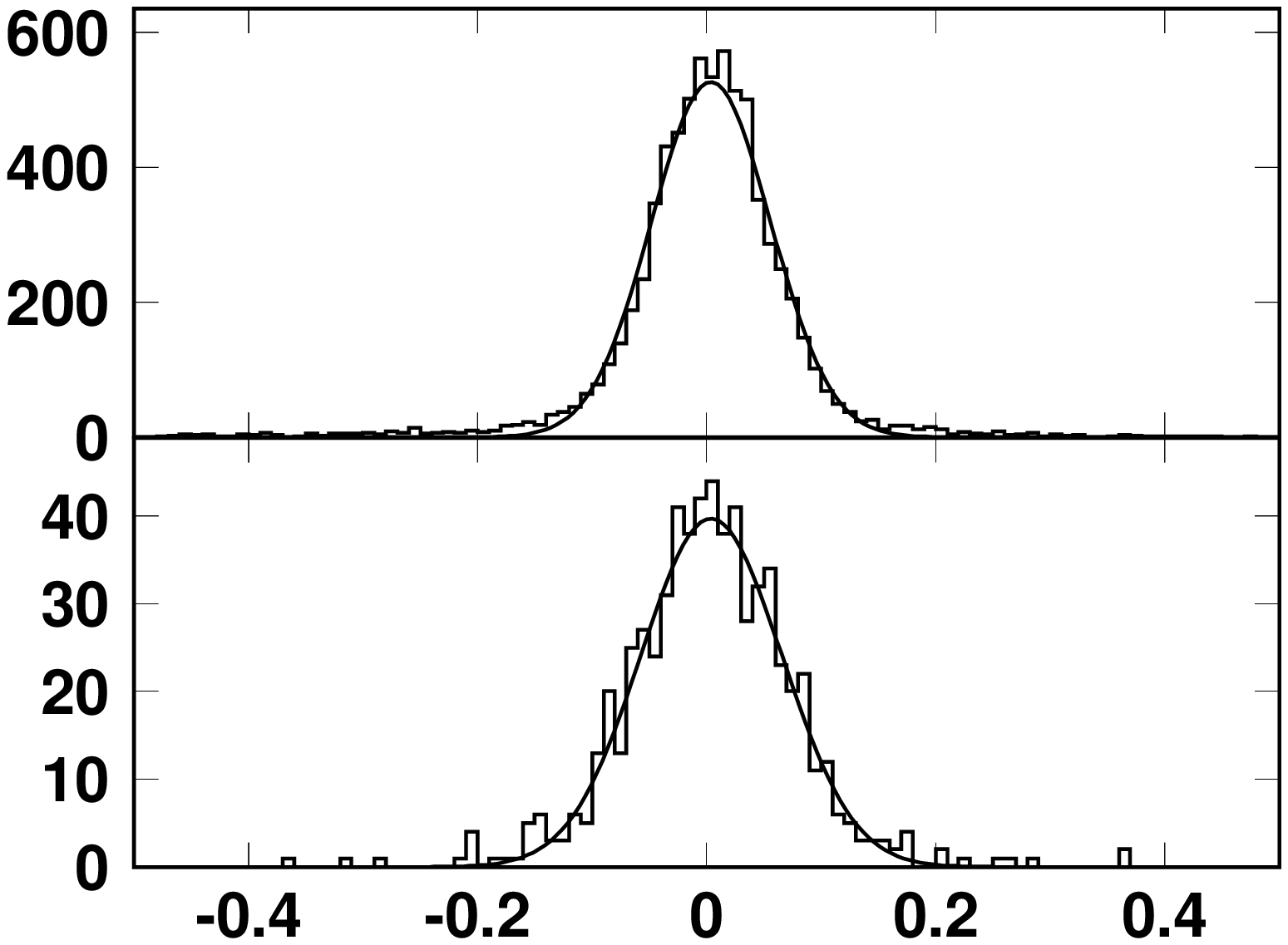}
\put(-130,10){$U_{miss}$ (GeV)}
\put(-250,80){\rotatebox{90}{Events/(0.010 GeV)}}
\put(-180,175){(a)}
\put(-180,100){(b)}
\caption{
The distributions of the $U_{miss}$ calculated for the Monte Carlo events
of (a) $D^+ \rightarrow \mu^+ \nu_{\mu}$ versus the tags of 
$D^- \rightarrow K^+\pi^-\pi^-$ and (b) $D^+ \rightarrow \mu^+ \nu_{\mu}$ versus 
the tags of $D^- \rightarrow K^+\pi^-\pi^-\pi^0$.
}
\label{umiss_mc}
\end{figure}

\begin{figure}[hbt]
\includegraphics[width=9.0cm,height=6.5cm]
{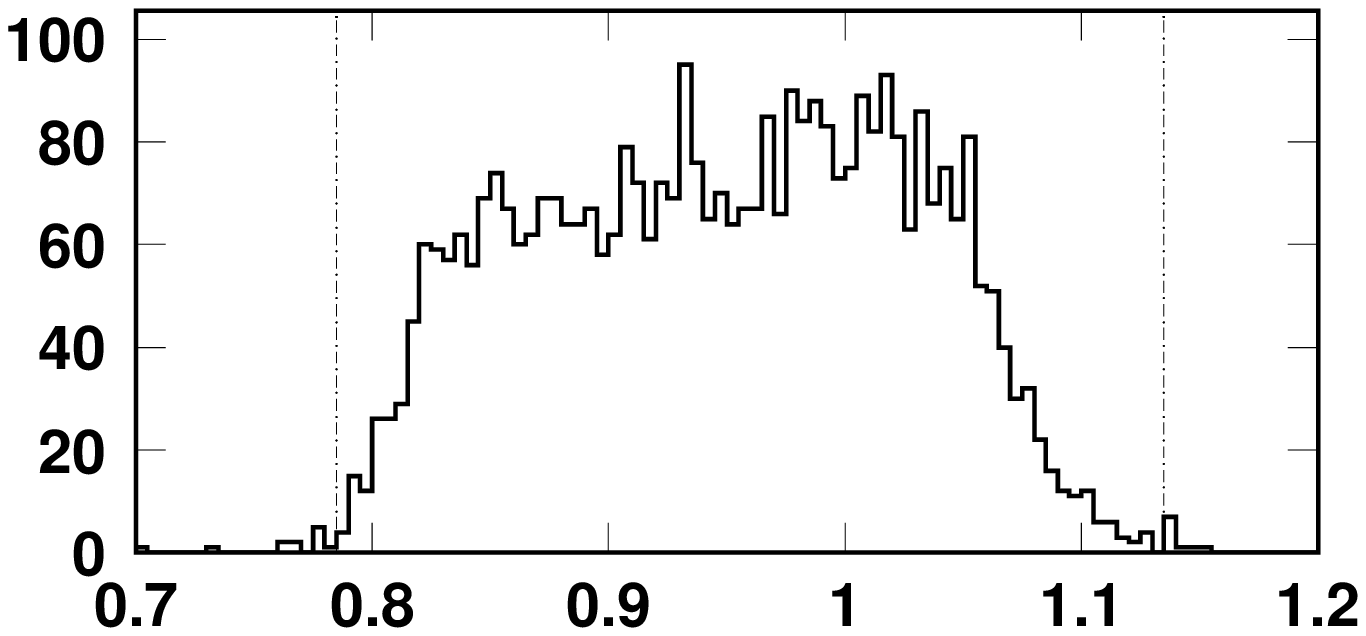}
\put(-150,15){$\mu^+$ momentum (GeV/$c$)}
\put(-255,60){\rotatebox{90}{Events/(0.005 GeV/$c$)}}
\caption{
The distribution of the reconstructed muon momentum selected from the
Monte Carlo events of $D^+ \rightarrow \mu^+\nu_{\mu}$; the interval
between the two dashed lines shows the region of the selected muon momentum in the
data analysis.
}
\label{momentum_mc}
\end{figure}

Candidate events for the decay $D^+ \rightarrow \mu^+\nu_{\mu}$
are selected from the surviving charged track in the system recoiling against the
singly tagged $D^-$ mesons. 
To select the $D^+ \rightarrow \mu^+ \nu_{\mu}$,
it is required that there be a single charged track
originating from the interaction region
in the recoil system of the $D^-$ tag and the charged track is identified
as a muon with charge opposite to the charge
of the tagged $D^-$.
For the candidate event, 
no extra good photon which is not used in the reconstruction
of the singly tagged $D^-$ meson
is allowed to be present in the event.
Since there is a missing neutrino
in the purely leptonic decay event, the event should
be characteristic with missing energy $E_{miss}$ and
missing momentum $P_{miss}$
which are carried by the neutrino.
For the purely leptonic decay event, the reconstructed
$U_{miss}$ which is defined as the difference between the $E_{miss}$ and
the $P_{miss}$ should be close to zero.
Fig.~\ref{umiss_mc}(a) shows the distribution of the $U_{miss}$
for the Monte Carlo events of $e^+e^- \rightarrow D^+D^-$, where
$D^- \rightarrow K^+\pi^-\pi^-$ and
$D^+ \rightarrow \mu^+\nu_{\mu}$.
Fig.~\ref{umiss_mc}(b) shows the same distribution of the
$U_{miss}$ for the Monte Carlo events of
$D^- \rightarrow K^+\pi^-\pi^-\pi^0$ and
$D^+ \rightarrow \mu^+\nu_{\mu}$.
The distribution of the reconstructed muon momentum selected from the
Monte Carlo events of $D^+ \rightarrow \mu^+\nu_{\mu}$ is
shown in Fig.~\ref{momentum_mc}, where the interval
between the two dashed lines is defined as the selected muon momentum region
for the events of $D^+ \rightarrow \mu^+ \nu_{\mu}$.

\begin{figure}[hbt]
\includegraphics[width=9.0cm,height=9.0cm]
{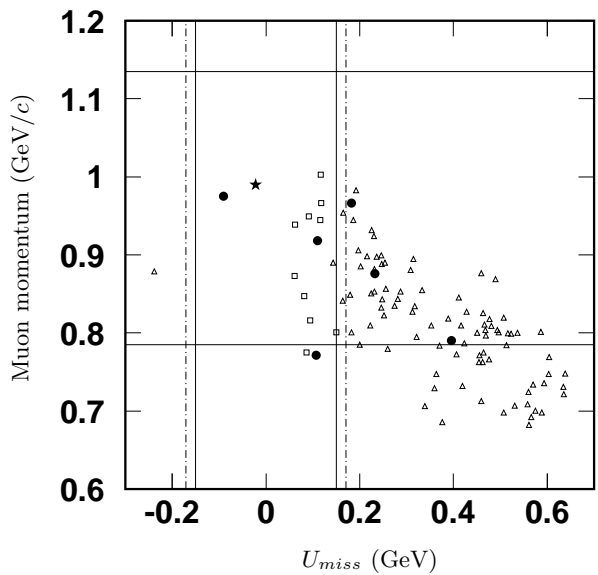}
\put(-150,10){$U_{miss}$ (GeV)}
\put(-260,80){\rotatebox{90}{Muon momentum (GeV/$c$)}}
\caption{
The scatter-plot of the muon momentum versus $U_{miss}$ recoiling
against the muon and the 
single tags ($mKm\pi$ combinations)
for the surviving purely leptonic
decay candidates, where
the dots $({\large \bullet})$ and the star $(\large {\star})$ are for the
single tag modes of
$K^+\pi^-\pi^-$ and $K^+\pi^-\pi^-\pi^0$,
respectively, those are from the data; the open squares 
and the open triangles 
are for the Monte Carlo background events from the Monte
Carlo sample which is 27 times larger than the data, see text.
}
\label{pmu_vs_umiss} 
\end{figure}

To select the purely
leptonic decay events from the singly tagged $D^-$ event sample,
it is required that the $U_{miss}$
of the candidate events should be
within  the $\pm 3\sigma_{U_{miss,i}}$ region for the single tag mode($i$),
where $\sigma_{U_{miss,i}}$ is the standard deviation of the $U_{miss,i}$
distribution obtained from the
Monte Carlo simulation for the event of 
$D^+ \rightarrow \mu^+\nu_{\mu}$ versus the single tag mode($i$) ($i=1$ is for
$K^+\pi^-\pi^-$; $i=2$ is for $K^0\pi^-$... and $i=9$ is for $\pi^+\pi^-\pi^-$
mode).
A further criterion requires that the candidate muon momentum
should be in the region from
0.785 to 1.135 GeV/$c$ as shown in Fig.~\ref{momentum_mc}.
Fig.~\ref{pmu_vs_umiss} shows the scatter-plot
of the momentum of the candidate muon versus the
$U_{miss}$, where
the dots and star are for the tag modes of
$K^+\pi^-\pi^-$ and $K^+\pi^-\pi^-\pi^0$,
respectively. 
The solid (dashed) vertical lines 
give the $\pm 3\sigma_{U_{miss}}$ interval for the single
tag mode of $D^- \rightarrow K^+\pi^-\pi^-$
($D^- \rightarrow K^+\pi^-\pi^- \pi^0$), 
while the two horizontal lines give the
selected momentum region for the $\mu^+$ 
from the purely leptonic decays of the $D^+$ mesons.
The regions surrounded by the lines are defined
as the selected signal regions for the two single tag modes.
There are 3 events 
within the signal regions for the two single tag modes. 

\begin{figure}[hbt]
\includegraphics[width=9.5cm,height=9.0cm]
{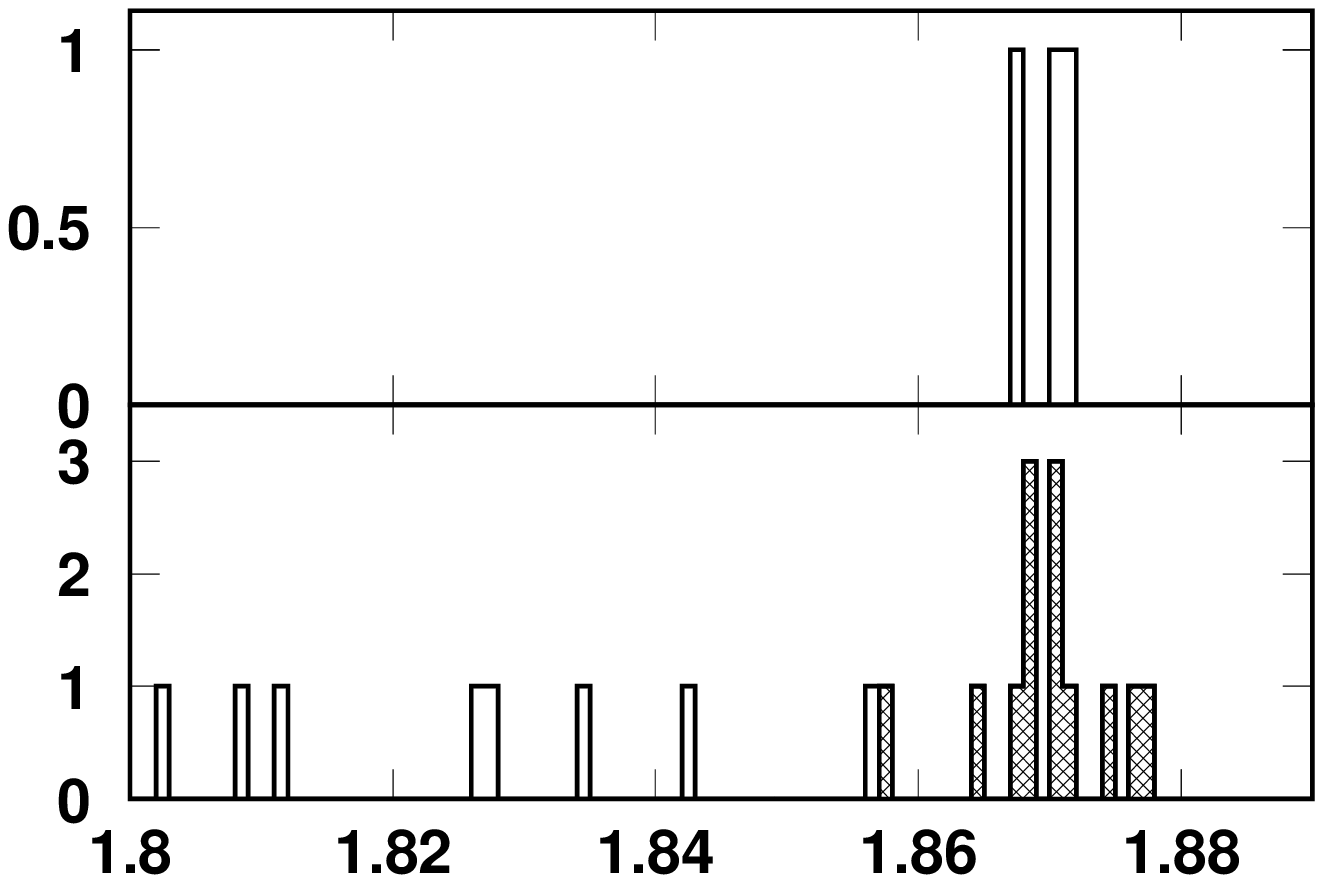}
\put(-170,10){Fitted Mass (GeV/$c^2$)}
\put(-275,85){\rotatebox{90}{Events/(0.001 GeV/$c^2$)}}
\put(-210,190){(a)}
\put(-210,105){(b)}
\caption{
The distributions of the $mKn\pi$ ($m=0$ or $1$ or $2$ and $n=1$ or $2$ or
$3$ or $4$) combinations for the events 
in which (a) the candidate events for $D^+ \rightarrow \mu^+\nu_{\mu}$
are found in the system recoiling against the tags ($mKn\pi$ combinations)
and (b) only one charged track is found in the system recoiling
against the tags and the events satisfy the kinematic requirements
for selecting the purely leptonic decay events,
but these events do not satisfy muon identification requirements.
}
\label{mass_tags_muv}
\end{figure}

Fig.~\ref{mass_tags_muv}(a) shows the distribution of the
fitted masses of the $mKn\pi$ combinations for the events which satisfy the
selection criteria.
The fitted masses of the 3 candidate events are
all in the well measured $D^-$ signal region in the mass spectrum.
Table~\ref{tbl_candites} summarizes the characteristics of the 3 events.
In each case, the measured mass of $D^-$ meson, the $U_{miss}$ of the event and
the momentum of the $\mu^+$ agree well with the expected values.
\begin{table}
\caption{Three candidates for $D^+$ purely leptonic decay.}
\label{tbl_candites}
\begin{center}
\begin{tabular}{cccc} \hline
 Event   & 1  & 2 & 3 \\
\hline
Tagging mode & $K^+\pi^-\pi^-$ & $K^+\pi^-\pi^-\pi^0$ & $K^+\pi^-\pi^-$\\
Fitted mass [MeV/$c^2$] & $1870.8$   &  $1876.9$  & $1871.4$ \\
Number of $\mu$ layer hits &  2              &   2   & 2 \\
$\mu^+$ momentum [GeV/$c$]   & 0.974   &  0.981  & 0.919 \\
$U_{miss}$  [GeV]    &  -0.093         &  -0.023 & 0.117 \\
Calculated momentum  &          &         &       \\
of neutrino [GeV/$c$] &  1.000  & 1.007  & 0.843  \\
\hline
\end{tabular}
\end{center}
\end{table}

\subsubsection{Background subtraction}

Some non-purely leptonic events from the $D^+$ decays
may also satisfy the selection criteria 
and are the background to the purely
leptonic decay events.
These background events must be subtracted.
The number of the background events is estimated by analyzing the
Monte Carlo sample which is 27 times larger than the data.
The Monte Carlo events are generated as
$e^+e^- \rightarrow D \overline D$ and the $D$ and $\bar D$
mesons are set to decay to all possible final states
according to the decay modes and the branching fractions
quoted from PDG~\cite{pdg} except the purely leptonic decay mode under study.
In Fig.~\ref{pmu_vs_umiss}, the open squares $({\tiny \Box})$ and
the open triangles $({\tiny \triangle})$ are for the 
background events from the
Monte Carlo sample, where the ${\tiny \Box}$ and
the ${\tiny \triangle}$ indicate that the background events are within and
outside of the $\pm 3\sigma_{U_{miss,i}}$ interval, respectively.
There are 9 background events satisfying the
selection criteria in the signal regions.
This number of the background events 
is then normalized to the corresponding data set.
A total of $0.33 \pm 0.11$ background events are estimated 
in the 3 candidates for $D^+ \rightarrow \mu^+\nu_{\mu}$.
After subtracting this number of the background events, 
$2.67\pm 1.74$ signal events
for $D^+ \rightarrow \mu^+\nu_{\mu}$ decay are retained.

The number of the background events can also be estimated from the 
number of the singly tagged $D^-$ events in which only one charged track
in the recoiling system is found and the charged track is not
identified as a muon.
Fig.~\ref{mass_tags_muv}(b) shows the distribution of the fitted masses 
of the $mKn\pi$ combinations from the events.
These events satisfy
the kinematic requirements for selecting the purely leptonic decay events, 
but the charged track in the recoiling system do not satisfy
the muon identification requirements.
There are 13 events in the $D^-$ signal regions as shown in the hatched
histogram.
The average probability of misidentifying a pion as a muon discussed
previously yields the number of the background events in the 3
purely leptonic decay candidates to be 
$0.31 \pm 0.09$,
which is consistent with $0.33 \pm 0.11$ estimated from the Monte Carlo
sample\footnote{
In response to the comment of the paper~\cite{cleoc_fd}, 
we here describe the background estimation in more detail.
We estimated the number of background events using two different methods.
In the first method, we estimated the number of background events using
the Monte Carlo events of $e^+e^- \rightarrow D\bar D$. 
The Monte Carlo sample of $D\bar D$ production and decays are generated
as $e^+e^- \rightarrow D^0\bar D^0, D^+D^-$, where the ratio of the neutral
over the total $D\bar D$ production cross section is set to be 0.58. Both
$D$ and $\bar D$ mesons decay to all possible modes 
(which of course include $D^+\rightarrow \pi^+\pi^0, \bar K^0 \pi^+, ...$)
according to the
branching fractions quoted from PDG~\cite{pdg}.
%
The decay channel $D^+\rightarrow \tau^+ \nu$ which is not currently listed 
in the PDG was added into the Monte Carlo sample to estimate the number of
background events. The branching fraction for $D^+\rightarrow \tau^+ \nu$
is estimated based on assuming decay constant $f_{D^+}=220$ MeV. 
In the second method, we use the data (see Fig. 6) to estimate the number of
background events. The number of background events obtained from the data
include all possible contribution from other $D$ decay channels
as well as the possible continuum backgrounds.}.

\section{Result}

\subsection{Monte Carlo Efficiency}

The efficiency for reconstruction of the purely leptonic decay
$D^+ \to \mu^+\nu_{\mu}$ after tagging the $D^-$ meson is estimated
by Monte Carlo simulation. A Monte Carlo study gives
that the efficiency is $\epsilon_{\mu^+\nu_{\mu}}=(41.7 \pm 1.1)\%$.

\subsection{Branching fraction}

To determine the purely leptonic decay branching fraction
and the decay constant based
on the observed numbers of the purely leptonic decay events
and the singly tagged $D^-$ mesons, we build a likelihood function $L$, which is the
product
of the Poissonian probability function for observation of the purely leptonic
decay events and the Gaussian function for the number of the singly tagged
$D^-$ mesons. Let $N_{\mu^+\nu_{\mu}}$ be the number of the observed
purely leptonic decay events ($N_{\mu^+\nu_{\mu}}=3.0 $),
$N_{tag}$ be the number of the singly
tagged $D^-$ mesons ($N_{tag}=5321 \pm 149 \pm 160$),
$n_{\mu^+\nu_{\mu}}$ and
$n_{tag}$ be the corresponding expected numbers of the events,
respectively.
The likelihood function is then given by
\begin{equation}
 L = P(n_{\mu^+\nu_{\mu}},
       N_{\mu^+\nu_{\mu}})G(n_{tag}, N_{tag}),
\end{equation}

\noindent
with abbreviation
$$P(n_{\mu^+\nu_{\mu}},N_{\mu^+\nu_{\mu}})=
  \frac {(n_{\mu^+\nu_{\mu}})^{N_{\mu^+\nu_{\mu}}}
        } {N_{\mu^+\nu_{\mu}} !} e^{-n_{\mu^+\nu_{\mu}}}
$$
\noindent
and
\begin{equation}
G(n_{tag}, N_{tag})=\frac
{1} {\sqrt{2\pi} \sigma_{N_{tag}} }
e^{-\frac{(N_{tag}-n_{tag})^2}{2 \sigma_{N_{tag}}} },
\end{equation}
\noindent
where the $\sigma_{N{tag}}$
is the systematic uncertainty in the number of the
singly tagged $D^-$ mesons ($\sigma_{N_{tag}}=160$).

\begin{figure}[hbt]
\includegraphics[width=9.5cm,height=7.0cm]
{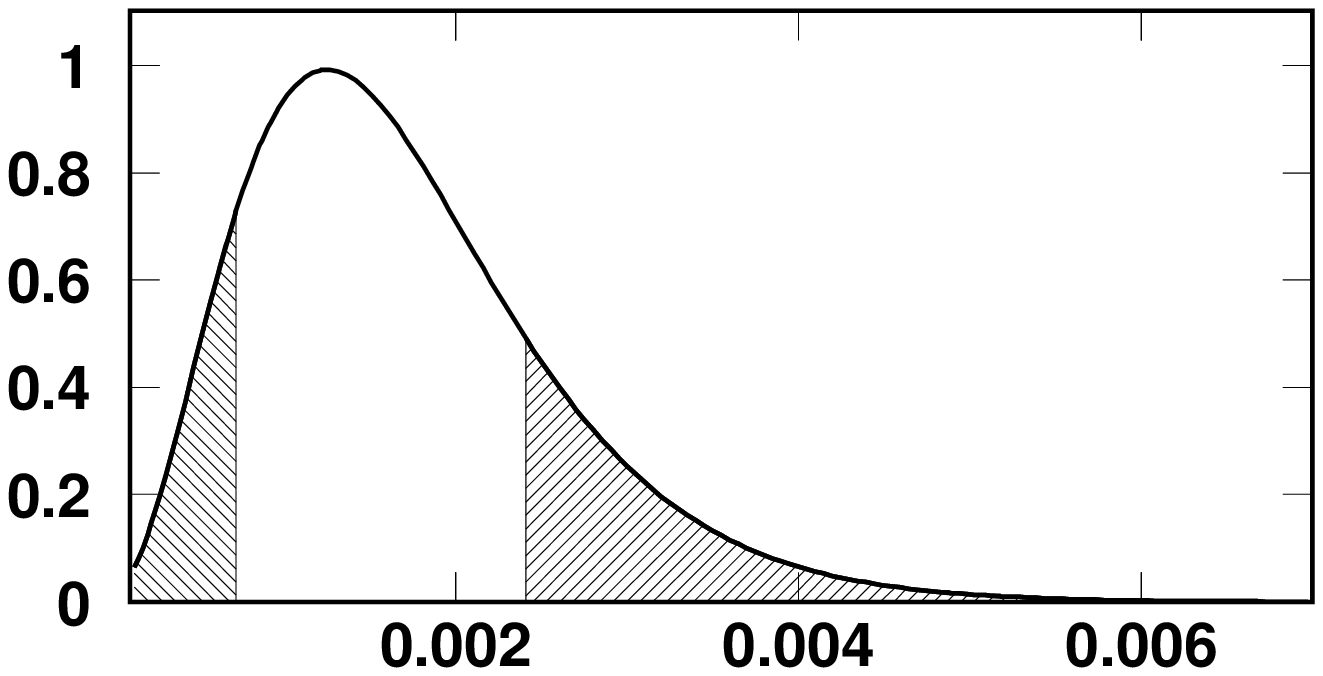}
\put(-170,10){$BF(D^+ \rightarrow \mu^+\nu_{\mu})$}
\put(-270,35){\rotatebox{90}{Normalized Likelihood $L/L_{max}$}}
\caption{The dependence of the normalized likelihood on the branching fraction
for the decay of $D^+ \rightarrow \mu^+\nu_{\mu}$.
}
\label{likehood_vs_bf}
\end{figure}

For the observed numbers of the purely leptonic decay events and the singly
tagged $D^-$ mesons, the expected number of the purely leptonic decay events
is given by
\begin{equation}
 n_{\mu^+\nu_{\mu}} = n_{tag} BF \epsilon_{\mu^+\nu_{\mu}} + n_b,
\end{equation}
\noindent
where $BF$ is the purely leptonic decay branching
fraction and $n_b$ is the number of the background events.
The value of the likelihood function is obtained 
by integrating over $N_{\mu^+\nu_{\mu}}$,
\begin{equation}
L(BF) = \int L(BF, N_{\mu^+\nu_{\mu}}) d{N_{\mu^+\nu_{\mu}}},~~~~
\end{equation}
\noindent
which is shown as a function of $BF$ in
Fig.~\ref{likehood_vs_bf}.
The maximum likelihood occurs at $BF=0.122\%$.
Integrating the function from the maximum position to $-1\sigma$
($+1\sigma$) value
corresponding to $68.3\%$ of the total area below (above) the peak position
yields the statistical error to be
$-0.053\%$ ($+0.111\%$).
The relative systematic uncertainty arises mainly from the uncertainties in
the $\mu^+$ identification ($\pm 5.0\%$),
tracking efficiency ($\pm 2.0\%$),
background subtraction ($\pm 5.6\%$) and
$U_{miss}$ cut ($\pm 1.0\%$).
Adding these uncertainties in quadrature gives
the total relative systematic uncertainty to be $\pm 7.8 \%$.
Finally, we obtain
$$BF(D^+ \rightarrow \mu^+\nu_{\mu})=(0.122^{+0.111}_{-0.053}\pm 0.010)\%.$$

\subsection{Decay constant $f_{D^+}$}

\begin{figure}[hbt]
\includegraphics[width=9.5cm,height=7.0cm]
{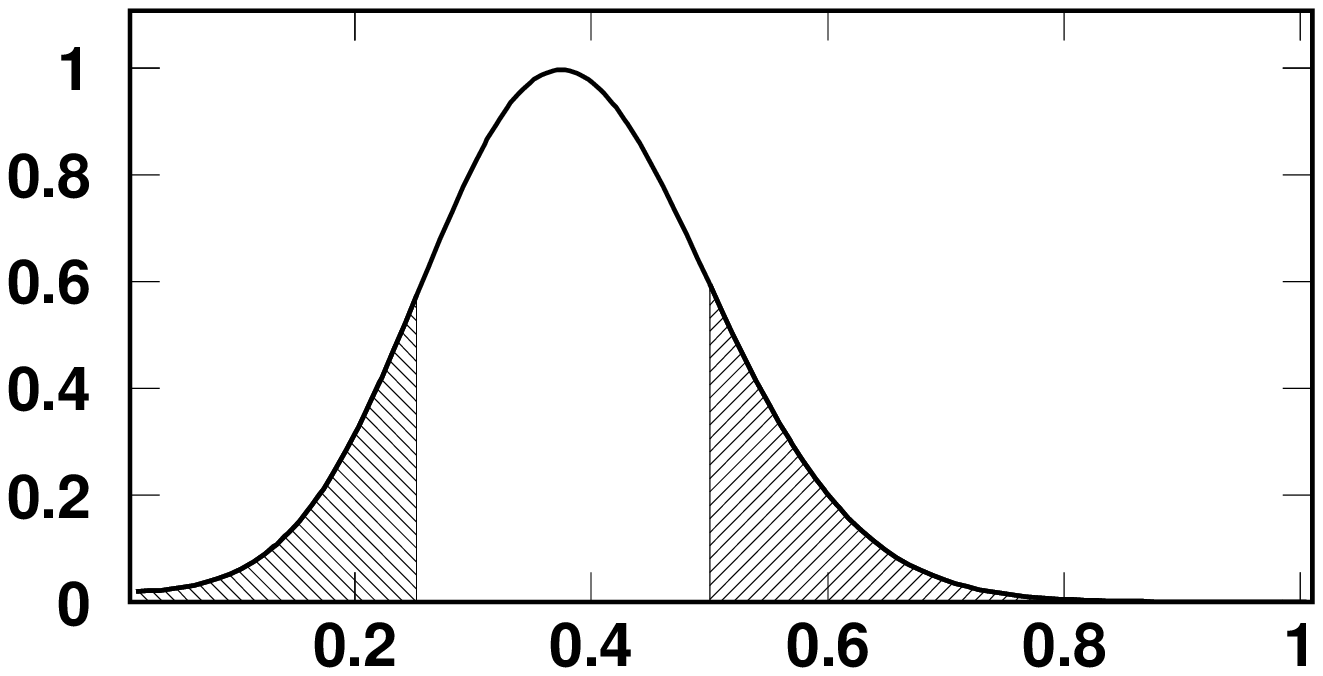}
\put(-170,10){$f_{D^+}$ (GeV)}
\put(-270,35){\rotatebox{90}{Normalized Likelihood $L/L_{max}$}}
\caption{The dependence of the normalized likelihood on the decay constant
$f_{D^+}$.
}
\label{likehood_vs_fd}
\end{figure}

The decay constant $f_{D^+}$ can  be obtained 
by inserting the mass of the muon, the mass of the $D^+$ meson, 
the CKM matrix element $|V_{cd}|$, the Fermi coupling constant $G_F$
and the lifetime of the $D^+$ meson~\cite{pdg}
into equation (1).
By substituting $BF(D^+ \rightarrow \mu^+\nu_{\mu})$ in terms of $f_{D^+}$
into equation (5), we obtain the relation of the likelihood function with
$f_{D^+}$,
\begin{equation}
L(f_{D^+}) = \int L(f_{D^+},N_{\mu^+\nu_{\mu}}) d{N_{\mu^+\nu_{\mu}}}.~~~~~
\end{equation}
\noindent
The most probable value of $f_{D^+}$ and its statistical error can be obtained
by integrating over $N_{\mu^+\nu_{\mu}}$. The dependence of the
likelihood function on the decay constant $f_{D^+}$ 
is shown in Fig.~\ref{likehood_vs_fd}.
Following the procedure
for extracting the $BF(D^+ \rightarrow \mu^+\nu_{\mu})$, we finally obtain
$$f_{D^+} = (371^{+129}_{-119}\pm 25)~~~\rm MeV,$$
\noindent
where the first error is statistical and the second systematic which arises
mainly from the uncertainties in 
the measured branching fraction ($\pm 4.1\%$), 
the CKM matrix element $|V_{cd}|$ ($\pm 5.4 \%$),
and the lifetime of the $D^+$ meson ($\pm 0.3\%$)~\cite{pdg}.
These uncertainties are added in quadrature to obtain the total systematic
error, which is $\pm 6.8\%$.

\section{Summary}

From the $5321 \pm 149 \pm 160$ singly tagged $D^-$ mesons, 
$2.67 \pm 1.74$ purely leptonic decay
events of $D^+ \rightarrow \mu^+ \nu_{\mu}$ are observed 
in the system recoiling against the singly tagged
$D^-$ mesons, resulting in the branching fraction of 
$BF(D^+ \rightarrow \mu^+\nu_{\mu}) =
(0.122^{+0.111}_{-0.053}\pm 0.010) \%$ and
the decay constant of $f_{D^+} = (371^{+129}_{-119}\pm 25)~\rm MeV$.
The measured values are independent of the $D^+D^-$ cross section and
luminosity. They do not require model-dependent assumptions. Thus, they are
absolute measurements. The central value of $f_{D^+}$ is consistent with
that measured using the BES-I detector~\cite{bes_fd}. The measured value of
$f_{D^+}$ is also consistent with most theoretical predictions, 
which are in the range from 90 to 360 MeV~\cite{prdct_fd}.

\vspace{5mm}

\begin{center}
{\small {\bf ACKNOWLEDGEMENTS}}
\end{center}
\par
\vspace{0.4cm}

   The BES Collaboration thanks the staff of BEPC for their hard efforts.
This work is supported in part by the National Natural Science Foundation
of China under contracts
Nos. 19991480,10225524,10225525, the Chinese Academy
of Sciences under contract No. KJ 95T-03, the 100 Talents Program of CAS
under Contract Nos. U-11, U-24, U-25, and the Knowledge Innovation Project
of CAS under Contract Nos. U-602, U-34(IHEP); by the
National Natural Science
Foundation of China under Contract No.10175060(USTC),and
No.10225522(Tsinghua University).

\end{document}